# Highly Uniform 300 mm Wafer-Scale Deposition of Single and Multilayered Chemically Derived Graphene Thin Films

Hisato Yamaguchi[#], Goki Eda[+], Cecilia Mattevi[+], HoKwon Kim[+], and Manish Chhowalla*,[+]

Department of Materials Science and Engineering, Rutgers University, 607 Taylor Road, Piscataway, NJ 08854, USA

[#]E-mail: hisaoty@rci.rutgers.edu

*E-mail: m.chhowalla@imperial.ac.uk

[+]Present Address: Department of Materials, Imperial College London, Exhibition Road, London SW7 2AZ, UK.

## ABSTRACT

The deposition of atomically thin highly uniform chemically derived graphene (CDG) films on 300 mm $SiO_2$/Si wafers is reported. We demonstrate that the very thin films can be lifted off to form uniform membranes than can be free-standing or transferred onto any substrate. Detailed maps of thickness using Raman spectroscopy and atomic force microscopy (AFM) height profiles reveal that the film thickness is very uniform and highly controllable, ranging from 1-2 layers up to 30 layers. After reduction using a variety of methods, the CDG films are transparent and electrically active with FET devices yielding exceptionally high mobilities of ~ 15 $cm^2$/Vs and sheet resistance of ~ 1 k$\Omega$/sq at ~ 70 % transparency.

Graphene[1] is emerging as a promising material that has the potential to revolutionize materials physics as well as much of present day electronics. While graphene could be viewed as the material for next generation of electronics, reports on making it technologically feasible for integration into devices are only recently beginning to emerge.[2-5] It is important to recognize that the extra-ordinary fundamental properties of a novel material do not necessarily translate into technological innovations and eventual implementation into applications. For example, it is useful to note that despite the investment of large amount of effort and resources over nearly two decades along with their attractive properties, single walled carbon nanotubes (SWNTs) have not been implemented into mainstream nano-electronics due to the absence of large area deposition and device integration due to lack of control over chirality and organization. Instead, large area or macro-electronics applications on flexible substrates requiring moderate performance devices where solution processed networks of SWNTs can be utilized have recently gained prominence.[6-7] Graphene will face similar challenges and thus it is important to tackle the obstacles of large area deposition and integration at an early stage of its development. In contrast to one dimensional nanomaterials, the synthesis of high quality graphene films on large area, CMOS compatible substrates will allow device fabrication using the well-known micro-electronics paradigm for thin film processing driven by continuing progress in lithography (top down approach). This represents a clear advantage over other materials that require novel approaches for their organization in dense arrays, capable of yielding device densities competitive with ULSI requirements.

There are two approaches to ultra-large area deposition of graphene. The first approach deals with condensation of high quality graphene with controllable layers using chemical vapor deposition (CVD) for nano-electronics where materials with extra-ordinary mobility values are required. The CVD approach requires high temperatures and single crystal substrates, although recent progress on growth and transfer suggests that graphene on a variety of substrates is feasible.[5,8-10] The second approach, the topic of this contribution, is the utilization of large area deposition of chemically derived graphene (CDG) from solution for electronics where extraordinary electrical properties are not required for high performance devices on flexible platforms. Devices such as sensors,[11] nanoelectromechanical systems (NEMS),[12] transparent conductors,[3-5,8,13-17] transistors,[1,4-5,9,17-19] and field emitters,[20] along with potential photonic applications[21-22] have been demonstrated with CDG. Virtually all such devices require peripheral CMOS based components for logic and operation. For example, existing tin oxide based sensors utilize lithographically processed heaters which enhance the sensitivity as well as refresh the sensing material. Thus, 300 mm (a size compatible with existing CMOS tools) wafer scale deposition of highly uniform CDG thin films with controllable number of layers is the first step towards technological implementation.

## RESULTS AND DISCUSSION
### Key Factors for CDG Thin Film Deposition and Transfer

We have achieved uniform 300 mm wafer scale deposition with controllable number of CDG layers by modified spin coating method. The flow chart of the deposition method summarizing the above is shown in Figure 1 (a). The key to uniform deposition on 300 mm wafers was to precisely determine the concentration and volume of CDG in methanol suspension. That is, if the suspension was too concentrated then aggregation of CGD flakes occurred, leading to thickness variations. If the concentration was too low then uniform and continuous coverage of the wafer surface could not be achieved. In addition to the precise concentration, the amount of suspension volume was also an important parameter. Another important factor was allowing sufficient time for the dropped suspension to spread over the wafer surface prior to rotation. With the determination of all the necessary parameters (i.e., amount and concentration of casting solution, spreading time), deposition was performed by spin coating and blowing nitrogen gas at the center region of the $SiO_2$ (300 nm)/Si wafer, as indicated in Figure 1 (a) (iv).[12] The deposition of CDG was readily confirmed *via* observation of rapid color change from purple to uniform light blue. The number of CDG layers was controlled by varying the spin coating speed. Single to bi-layer films were achieved at 8000 rpm, 4-5 layers at 6000 rpm, and 7-8 layers at 4000 rpm (Supplementary Information). Thicker films could be achieved by decreasing the rotation speed to 2000 rpm or by repeating the deposition multiple times. Uniform thin films were deposited with close to 100 % reproducibility.

A schematic of the lift-off and transfer procedure of the thin films to obtain free-standing membranes is also shown in Figure 1 (a). The successful transfer of one dimensional nanostructures has been demonstrated over reasonably large areas.[26-28] The advantage of the method reported here is that it can transfer very uniform CDG thin films on 300 mm wafers with yield of 100 %. This procedure leads to uniform lift-off of membranes consisting of atomically thin CDG films with poly meta methyl acrylate (PMMA) support, which can be readily handled as bulk film. Pure CDG membranes can be transferred onto any substrate by scooping or remain free-standing after PMMA removal using solvents such as acetone. Furthermore, we did not observe cracks on our transferred films, which are reported to be present in the transferred CVD-grown graphene films.[29] The ability to fabricate free-standing CDG films indicates that the individual flakes within the thin films are well adhered and form a continuous network, which allows the membranes to maintain structural integrity even when the support PMMA is washed away. Photographs of 150 mm and 300 mm wafers (6 and 12 inches) with atomically thin layers of uniform CDG and the corresponding free-standing membranes are shown in Figure 1 (b) and (c), respectively.

**AFM, Optical Microscope, and Camera-Taken Images**

A 300 mm membrane transferred onto PET film is shown in Figure 2 (a). The transferred membrane is flexible and transparent. Typical optical microscope image of deposited graphene membrane without PMMA is shown in Figure 2 (b). Scale bar indicates 50 μm. The membrane is continuous with overlapping flakes. Flake sizes generally range from 20-30 μm, consistent with our previous observations.[19] The large size of the CDG flakes synthesized in our laboratory is an important factor in successful deposition of ultra-large area CDG films. AFM images of the thin films show CDG flakes laying flat on the substrate with almost no visible wrinkles (Figure 2 (c)). Scale bar indicates 1 μm. We noticed much lower density of wrinkles on the CDG sheets compared to films deposited by vacuum filtration, especially for very thin membranes. This can be explained by combination of rotation force applied horizontal to substrate during spin coating and rapid evaporation of solvent to "freeze" flakes onto the substrate. Height profile of a cross section indicated by the green dashed line is included in the AFM image. Step-like height difference of ~ 0.67 nm can be seen at the region corresponding to the edges of the flakes. It can also be seen that CDG sheets are also present in the region between the flakes. Our thinnest membranes consist of thickness as small as ~ 0.7 nm (Supplementary Information), which shows that single layer CDG membranes over 300 mm can be achieved with our method.

**Raman and XPS**

Raman spectroscopy was performed to confirm the thickness of the thin films by measuring the number of reduced CDG layers. Raman spectra of reduced CDG membrane were similar to our previous results[4] in that it showed the presence of the usual D, G, and 2D peaks. Presence of D peak, which is absent in mechanically cleaved graphene,[30] indicates the presence of structural imperfections induced by the attachment of oxygen functional groups on the carbon basal plane. The intensity of the 2D peak with respect to the D and G peaks is small due to disorder but the shift and shape of this peak can be used for analyzing the number of layers in reduced CDG. We have performed similar analysis to our previous work[4] to determine the number layers in reduced CDG thin films. A typical Raman map over 48 μm x 24 μm spatial region is shown in Figure 2 (d). The result shows that the thin films deposited consist of mostly single to bi-layer, which is consistent with AFM image in Figure 2 (c).

For electronic applications of CDG thin films, reduction is essential because as-deposited GO is insulating.[4,13] We have attempted a number of methods to reduce CDG thin films. Specifically, we have reduced by dipping CDG flakes in hydrazine,[18] exposure to hydrazine and dimethylhydrazine vapor,[4,13-14] electrochemical reduction,[31] thermal annealing[3-4,13,15] and a combination of these techniques.[4,13] We have found that to a large extent, all reduction treatments lead to similar results. Figure 2 (e) shows C1s peak of X-ray Photoelectron Spectroscopy (XPS) spectrum on CDG thin films reduced *via* thermal annealing at 1100 °C. The most important feature of XPS was that oxygen content was found to have been reduced to ∼8 at. % from ∼39 at. %[32] for as-synthesized CDG. With the aid of peak fitting program, the peaks were deconvoluted into: C=C/C-C in aromatic rings (284.6 eV), C-O (286.1 eV), C=O (287.5 eV), C=O(OH) (289.2 eV), and π- π π * satellite peak (290.6 eV).[33] These assignments are in good agreement with our previous work.[32]

**Opto-Electronic and Transport Properties**

Transmittance and sheet resistance of reduced CDG thin films as a function of thickness are shown in Figure 3. Transmittance of ~ 96 % was measured for single and bi-layered films. Lowest sheet resistance of 600 Ω/sq, albeit at a low transparency of ~ 40 %, was achieved for reduced CDG films with thickness of ~ 30 nm. The best opto-electronics properties were obtained for lower thicknesses with a thin film of 5 nm exhibiting sheet resistance of ~ 2.4 kΩ/sq at a transmittance of ~ 81 % while 15 nm thin films exhibited transmittance of ~ 70 % at a sheet resistance of ~ 1 kΩ/sq. A recent report on the high temperature CVD growth and subsequent transfer of graphene films demonstrated better opto-electronic values than for the CDG thin films reported here.[29]

Typical transport measurement results for reduced CDG are shown in Figure 4. Conventional electron-beam lithography was used to define electrodes over deposited membrane. Cr/Au (5/30 nm) were thermally evaporated followed by lift-off of the mask. The channel length between source (S) and drain (D) was 20 μm. The "switching" behavior was measured by two-terminal configurations with silicon substrate used to apply gate bias ($V_g$). Measurements were made in vacuum, and in ambient conditions at room temperature. Conductance modulation by varying gate bias was observed in all measured devices, which showed ambipolar field effect similar to that of the previous study.[4] The device is unintentionally doped by holes most probably due to adsorbed oxygen and water molecules. The neutrality point is generally observed for gate bias of around + 50 V that shifted to almost 0 V when measurements were performed in vacuum. The on/off ratio of the devices was measured to be ~3 and thus is far from a practical switching device but the mobility values reported here can be used to define the quality of reduction treatment and also for comparison with other reports in the literature. The best mobility we observed was 10 - 15 cm$^2$/Vs for the hole branch of the transfer characteristics, with the electron mobilities being slightly lower.

## CONCLUSIONS

In summary, we have deposited uniform transparent and conducting CDG thin films with control over the number of layers. The lowest sheet resistance we achieved was 600 Ω/sq with field effect mobility values > 10 cm$^2$/Vs on CMOS compatible 300 mm SiO$_2$/Si wafers. The deposition over ultra-large areas is highly uniform and reproducible, yielding films ranging in thickness from single monolayer to several layers. We also demonstrate a simple method to obtain large free-standing membranes of CDG thin films using a simple lift-off method. The ability to obtain high mobility solution processed graphene thin films on wafers that are compatible with existing CMOS tools should provide a pathway for technological implementation of CDG thin film in applications such as sensors, NEMS and analog devices.

## METHODS
### Deposition of CDG Films

The precise concentration, volume, and spreading time for achieving uniform deposition for 300mm CDG thin film *via* modified spin coating method was 0.4 mg/ml, 60 ml, and 30 min., respectively. CGD was fabricated by oxidizing graphite to obtain graphene oxide (GO) using the modified Hummers method, details of which are provided elsewhere.[23] Prior to casting the suspensions, the wafers were dipped into 50 wt% potassium hydroxide (KOH) solution for 15 min. to enhance the hydrophilicity of the surface. Obvious decrease in contact angle upon casting the solution was observed after the treatment, enabling rapid spread of the suspension over the entire substrate area. The role of the surface treatment was essential for further enhancement in uniform spreading of CDG flakes. After the treatment, substrates were rinsed in water and dried.

After casting the GO solution onto center of the substrates, time was allowed prior to rotation (Spreading time). When rotation started (Spin coater, Model P6700, Specialty Coating Systems (SCS) Inc.), nitrogen gas was blown at center region of the substrates to accelerate the vaporization of the solvent.[12] Without the nitrogen gas blow, the deposited films were not uniform and continuous. When all the solvent was vaporized, CDG films were deposited on the substrates.

**Transfer of Deposited CDG Films**

For transfer, PMMA was casted directly onto deposited CDG films, and spin coated for 500 rpm for 5 sec. followed by 4000 rpm for 60 sec. The PMMA coated CDG films were post-baked for 2 hours at 170 $^o$C to enhance adhesion between PMMA supports. The films was then submerged (for 3hrs) in sodium hydroxide (NaOH, 1M) to remove the PMMA/CDG membranes from the substrate.[24-25] The membranes were dried for few hours and rinsed in acetone for 30 min. to remove the PMMA supports. We observed no significant change in the thickness of deposited films and transferred films, which is consistent with Reina *et al.*[24] showing successful transfer down to a single flake.

**Reduction**

*Directly in Hydrazine Anhydrate*

Dried GO was prepared for reduction directly in hydrazine anhydrate (Caution: hydrazine anhydrate is extremely toxic and should be handled in a glove box). GO solution was prepared in a petri dish and placed in a vacuum desiccator with phosphorus pentoxide for a week. When the GO was dry, it was directly mixed with hydrazine anhydrate in a glove box. After several hours, GO is reduced and dispersed in the solution.

*Hydrazine Vapor*

For reduction by hydrazine and dimethylhydrazine vapor (Caution: hydrazine and dimethylhydrazine are extremely toxic and should be handled with an extra care), few ml of the 1 vol % aqueous solution was prepared in a small glass petri dish. The petri dish with the solution was then placed in a larger glass petri dish with the samples inside, and loosely sealed. The petri dish was heated on a hot plate to 90 $^{o}$C for reduction. Reduction time was typically 2 hours.

*Electrochemical*

CDG films were deposited on ITO substrates for electrochemical reduction. Kapton tape was used to mask regions for electrode contacts. CDG films on ITO substrates were placed in a beaker filled with sulfuric acid ($H_2SO_4$) (1M), which acts as an electrolyte. Glassy carbon rod ($\phi$=1mm) was gently pressed onto ITO substrate where Kapton mask was used, and Ag rod ($\phi$=1mm) was placed in electrolyte as a counter electrode. Voltage ranging from 1.0-2.0 V was applied between the electrodes for reduction. About 20 sec. after voltage is applied, rapid change of the film color was observed, indicating reduction is taking place. Reduction time was about 5 min.

*Thermal*

GO films were annealed in a conventional furnace back filled with Ar/$H_2$ gas mixture (90 % Ar and 10% $H_2$). Annealing time was 15 min.

## ACKNOWLEDGEMENT

The authors would like to acknowledge J. T. Robinson of Naval Research Laboratory for his helpful suggestions at the initial stage of this work. We also acknowledge Y. Yamaguchi for help with camera-taken images. This work was funded by the National Science Foundation CAREER Award (ECS 0543867).

**REFERENCES**

1. Novoselov, K.S.; Geim, A.K.; Morozov, S.V.; Jiang, D.; Zhang, Y.; Dubonos, S.V.; Grigorieva, I.V.; Firsov, A.A. Electric Field Effect in Atomically Thin Carbon Films. *Science* **2004**, 306, 666-669.
2. Stankovich. S.; Dikin, D.A.; Dommett, G.H.B.; Kohlhaas, K.M.; Zimney, E.J.; Stach, E.A.; Piner, R.D.; Nguyen, S.T.; Ruoff, R.S. Graphene-Based Composite Materials. *Nature* **2006**, 442, 282-286.
3. Li, X.; Zhang, G.; Bai, X.; Sun, X.; Wang, X.; Wang, E.; Dai, H. Highly Conducting Graphene Sheets and Langmuir-Blodgett Films. *Nature Nanotech.* **2008**, 3, 538-542.
4. Eda, G.; Fanchini, G.; Chhowalla, M. Large-Area Ultrathin Films of Reduced Graphene Oxide as a Transparent and Flexible Electronic Material. *Nature Nanotech.* **2008**, 3, 270-274.
5. Reina, A.; Jia, X.; Ho, J.; Nezich, D.; Son, H.; Bulovic, V.; Dresselhaus M.S.; Kong J. Large Area, Few-Layer Graphene Films on Arbitrary Substrates by Chemical Vapor Deposition. *Nano Lett.* **2009,** 9, 30-35.
6. Wu, Z.; Chen, Z.; Du, X.; Logan, J.M.; Sippel, J.; Nikolou, M.; Kamaras, K.; Reynolds, J.R.; Tanner, D.B.; Hebard, A.F.; Rinzler, A.G. Transparent, Conductive Carbon Nanotube Films. *Science* **2004**, 305, 1273-1276.
7. LeMieux, M.C.; Roberts, M.; Barman, S.; Jin, Y.W.; Kim, J.M.; Bao, Z.; Self-Sorted, Aligned Nanotube Networks for Thin-Film Transistors. *Science*, **2008**, 321, 101-104.
8. Kim, K.S.; Zhao, Y.; Jang, H.; Lee, S.Y.; Kim, J.M.; Kim, K.S.; Ahn, J.-H.; Kim, P.; Choi, J.-Y.; Hong, B.H. Large-Scale Pattern Growth of Graphene Films for Stretchable Transparent Electrodes. *Nature* **2009**, 457, 706-710.
9. Li, X.; Cai, W.; An, J.; Kim, S.; Nah, J.; Yang, D.; Piner, R.; Velamakanni, A.; Jung, I.; Tutuc, E.; Banerjee, S.K.; Colombo, L.; Ruoff, R.S. Large-Area Synthesis of High-Quality and Uniform Graphene Films on Copper Foils. *Science* **2009**, 324, 1312-1314.
10. Feng, X.; Marcon, V.; Pisula, W.; Hansen, M.R.; Kirkpatrick, J.; Grozema, F.; Andrienko, D.; Kremer, K.; Müllen, K. Towards High Charge-Carrier Mobilities by Rational Design of the Shape and Periphery of Discotics. *Nature Mater.* **2009**, 8, 421-426.
11. Robinson, J.T.; Perkins, F.K.; Snow, E.S.; Wei, Z.; Sheehan, P.E. Reduced Graphene Oxide Molecular Sensors. *Nano Lett.* **2008**, 8, 3137-3140.
12. Robinson, J.T.; Zalalutdinov, M.; Baldwin, J.W.; Snow, E.S.; Wei, Z.; Sheehan, P.; Houston, B.H. Wafer-scale Reduced Graphene Oxide Films for Nanomechanical Devices. *Nano Lett.* **2008**, 8, 3441–3445.
13. Becerril, H.A.; Mao, J.; Liu, Z.; Stoletenberg, R.M.; Bao, Z.; Chen, Y. Evaluation of Solution-Processed Reduced Graphene Oxide Films as Transparent Conductors. *ACS Nano* **2008**, 2, 463-470.
14. Watcharotone, S.; Dikin, D.A.; Stankovich, S.; Piner, R.; Jung, I.; Dommett, G.H.B.; Evmenenko,

G.; Wu, S.-E.; Chen, S.-F.; Liu, C.-P.; Nguyen, S.T.; Ruoff, R.S. Graphene−Silica Composite Thin Films as Transparent Conductors. *Nano Lett.* **2007**, 7, 1888-1892.

15. Wang, X.; Zhi, L.; Müllen, K.; Transparent, Conductive Graphene Electrodes for Dye-Sensitized Solar Cells. *Nano Lett.* **2008**, 8, 323-327.

16. Eda, G.; Lin, Y.-Y.; Miller, S.; Chen, C.-W.; Su, W.-F.;Chhowalla, M. Transparent and Conducting Electrodes for Organic Electronics from Reduced Graphene Oxide. *Appl. Phys. Lett.* **2008**, 92, 233305.

17. Tung, V.C; Chen, L.-M.; Allen, M.J.; Wassei, J.K.; Nelson, R.B.; Yang, Y.; Low-Temperature Solution Processing of Graphene-Carbon Nanotube Hybrid Materials for High-Performance Transparent Conductors. *Nano Lett.* **2009**,

18. Eda, G.; Chhowalla, M. Graphene-Based Composite Thin Films for Electronics. *Nano Lett.* **2009**, 9, 814–818.

19. Tung, V.C.; Allen, M.J.; Yang, Y.; Kaner, R.B. High-Throughput Solution Processing of Large-Scale Graphene. *Nature Nanotech.* **2009**, 4, 25-29.

20. Eda, G.; Unalan, H.E.; Rupesinghe, N.; Amaratunga, G.A.J.; Chhowalla, M. Field Emission from Graphene Based Composite Thin Films . *Appl. Phys. Lett.* **2008**, 93, 233502.

21. Luo, Z.; Vora, P.M.; Mele, E.J.; Johnson, A.T.C.; Kikkawa, J. M. Photoluminescence and Band Gap Modulation in Graphene Oxide. *Appl. Phys. Lett.* **2009**, 94, 111909.

22. Eda, G.; Lin, Y.-Y.; Mattevi, C.; Yamaguchi, H.; Chen, H.-A.; Chen, I.-S.; Chen, C.-W.; Chhowalla, M. Blue Photoluminescence from Chemically Derived Graphene. *Adv. Mater.* **2009**, DOI: 10.1002/adma.200901996.

23. Hirata, M.; Gotou, T.; Horiuchi, S.; Fujiwara, M.; Ohba, M. Thin-Film Particles of Graphite Oxide 1: High-Yield Synthesis and Flexibility of the Particles. *Carbon* **2004**, 42, 2929-2937.

24. Reina, A.; Son, H.; Jiao, L.; Fan, B.; Dresselhaus, M.S.; Liu, Z.; Kong, J. Transferring and Identification of Single- and Few-Layer Graphene on Arbitrary Substrates. *J. Phys. Chem. C* **2008**, 112, 17741-17744.

25. Jiao, L.; Fan, B.; Xian, X.; Wu, Z.; Zhang, J.; Liu, Z. Creation of Nanostructures with Poly(methyl methacrylate)-Mediated Nanotransfer Printing. *J. Am. Chem. Soc.* **2008**, 130, 12612-12613.

26. Meitl, M.A.; Zhou, Y.; Gaur, A.; Jeon, S.; Usrey, M.L.; Strano, M.S.; Rogers, J.A. Solution Casting and Transfer Printing Single-Walled Carbon Nanotube Films. *Nano Lett.* **2004**, 4, 1643-1647.

27. Hines, D.R.; Mezhenny, S.; Breban, M.; Williams, E.D.; Ballarotto, V.W.; Esen, G.; Southard, A.; Fuhrer, M.S. Nanotransfer Printing of Organic and Carbon Nanotube Thin-Film Transistors on Plastic Substrates. *Appl. Phys. Lett.* **2005**, 86, 163101.

28. Chai, Y.; Gong, J.; Zhang, K.; Chan, P.C.H.; Yuen, M.M.F. Flexible Transfer of Aligned Carbon

Nanotube Films for Integration at Lower Temperature. *Nanotechnology* **2007**, 18, 355709.
29. Li, X.; Yanwu Zhu, Y.; Cai, W.; Borysiak, M.; Han, B.; Chen, D.; Piner, R.D.; Colombo, L.; Ruoff, R.S. Transfer of Large-Area Graphene Films for High-Performance Transparent Conductive Electrodes. *Nano Lett.* **2009**, DOI: 10.1021/nl902623y.
30. Ferrari, A.C.; Meyer, J.C.; Scardaci, V.; Casiraghi, C.; Lazzeri, M.; Mauri, F.; Piscanec, S.; Jiang, D.; Novoselov, K.S.; Roth, S.; Geim, A.K. Raman Spectrum of Graphene and Graphene Layers. *Phys. Rev. Lett.* **2006**, 97, 187401/1-187401/4.
31. Zhou, M.; Wang, Y.; Zhai, Y.; Zhai, J.; Ren, W.; Wang, F.; Dong, S. Controlled Synthesis of Large-Area and Patterned Electrochemically Reduced Graphene Oxide Films. *Chem. Eur. J.* **2009**, 15, 6116-6120.
32. Mattevi, C.; Eda, G.; Agnoli, S.; Miller, S.; Mkhoyan, K.A.; Celik, O.; Granozzi, G.; Garfunkel, E.; Chhowalla, M. Evolution of Electrical, Chemical, and Structural Properties of Transparent and Conducting Chemically Derived Graphene Thin Films. *Adv. Funct. Mater.* **2009**, 19, 1-7.
33. Hontoria-Lucas, C.; Lopez-Peinando, A.J.; Lopez-Gonzalez, J.D.D.; Rojas-Cervantes, M.L.; Martin-Aranda, R.M. Study of Oxygen-Containing Groups in a Series of Graphite Oxide: Physical and Chemical Characterization. *Carbon* **1995**, 33, 1585-1592.

**FIGUR CAPTIONS**

**Figure 1 (a)**
The complete flow chart of the deposition procedures for 300 mm CDG films. (iv) shows deposition performed by spin coating and blowing nitrogen gas at the center region of the $SiO_2$/Si wafer.

**Figure 1 (b), (c)**
Photographs of (b) 150 mm (6 inch) and (c) 300 mm (12 inch) wafers with atomically thin layers of uniform CDG and the corresponding free-standing membranes.

**Figure 2 (a)**
A 300 mm membrane transferred onto PET film. The transferred membrane is flexible and transparent.

**Figure 2 (b), (c)**
Typical optical microscope image (scale bar indicates 50 μm) and AFM image (scale bar indicates 1 μm) of deposited graphene membrane without PMMA, respectively.

**Figure 2 (d)**

A typical Raman map over 48 μm x 24 μm spatial region, which shows that the deposited film consists of mostly single to bi-layer.

**Figure 2 (e)**

C1s peak of XPS spectra on CDG thin films reduced *via* thermal annealing at 1100 °C. Oxygen content was found to have been reduced to ∼8 at. %.

**Figure 3**

Transmittance and sheet resistance of reduced CDG thin films as a function of thickness. Transmittance of ~ 96 % was measured for single and bi-layered films. Lowest sheet resistance achieved was 600 Ω/sq.

**Figure 4**

Typical transport measurement results for reduced CDG membrane performed in vacuum and air. Inset shows an optical microscope image of the actual device. The channel length between source (S) and drain (D) is 20 μm.

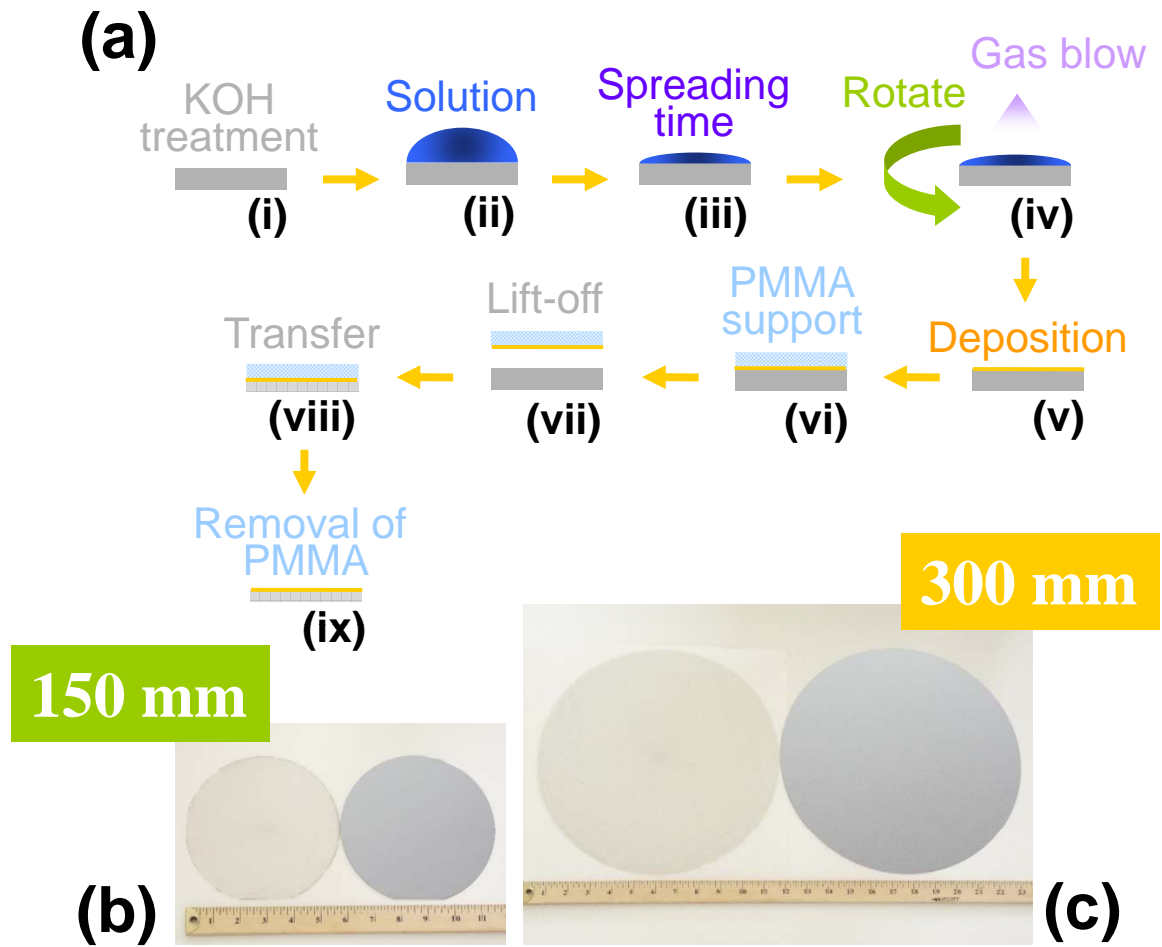

**Figure 1** H.Yamaguchi *et al.*

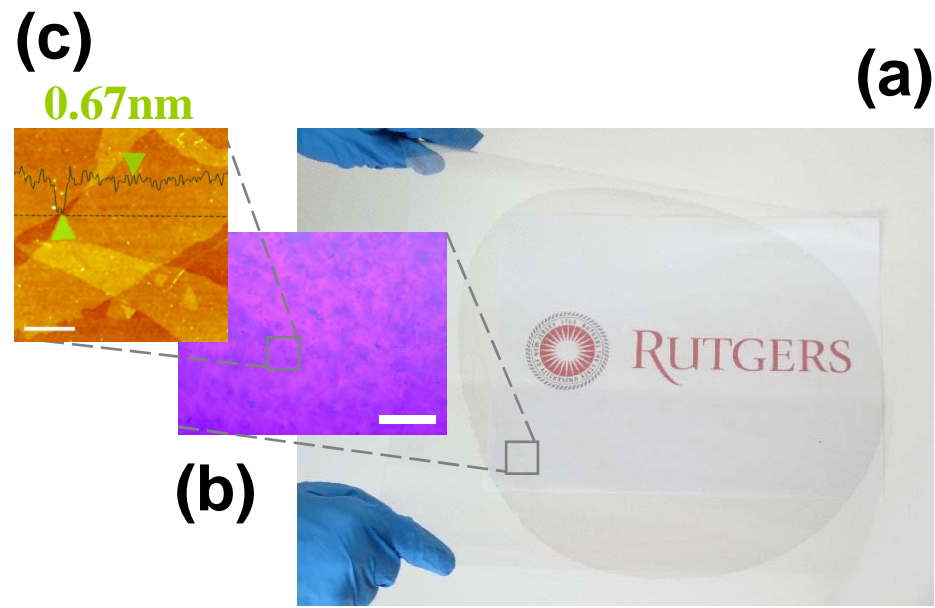

**Figure 2** H.Yamaguchi *et al.*

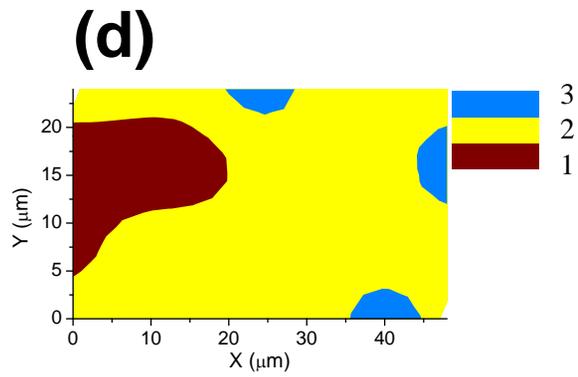 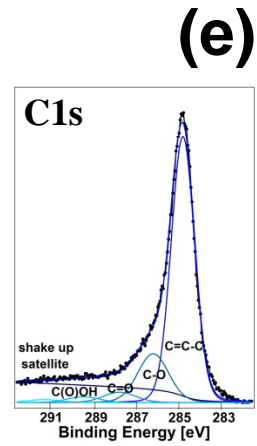

Figure 2 H.Yamaguchi *et al.*

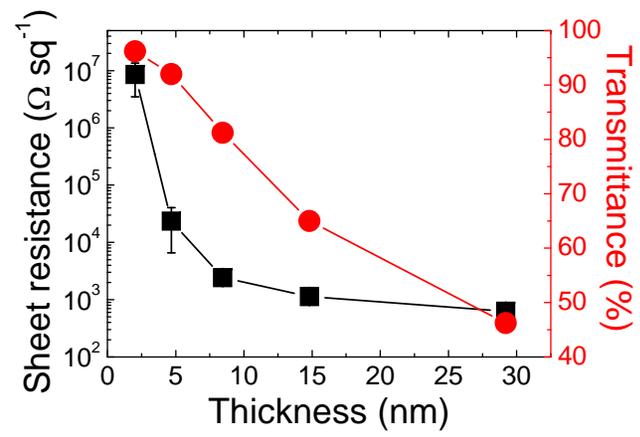

**Figure 3** H. Yamaguchi *et al.*

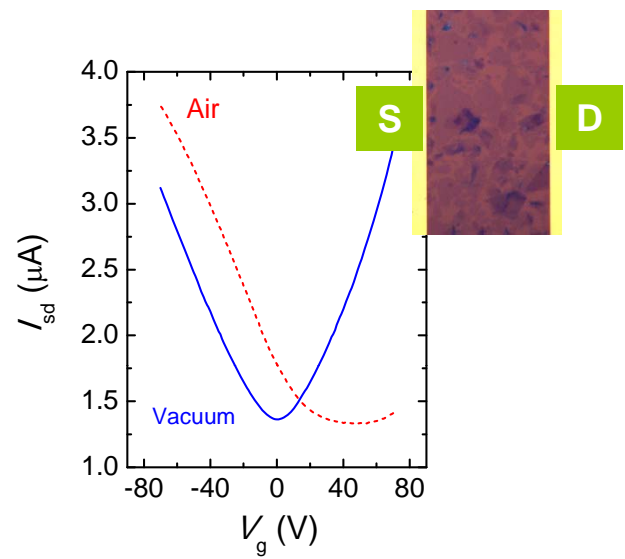

**Figure 4** H.Yamaguchi *et al.*